\documentclass[10pt,twocolumn]{article} 
\usepackage{simpleConference}
\usepackage{times}
\usepackage{graphicx}
\usepackage{amssymb}
\usepackage{url,hyperref}
\usepackage{tikz}
\usetikzlibrary{fit,positioning} % needed for some reason..
\usepackage{bm}					% bold mathematics
\usepackage{algorithmic}
\usepackage{textcomp}
\newcommand{\gmcsize}{5mm}
\newcommand{\gmcoffset}{15mm}
\def\BibTeX{{\rm B\kern-.05em{\sc i\kern-.025em b}\kern-.08em
    T\kern-.1667em\lower.7ex\hbox{E}\kern-.125emX}}

\begin{document}

\title{Using Topic Models to Mine Everyday Object Usage Routines Through Connected IoT Sensors}

\author{
  Yanxia Zhang\\
  Delft University of Technology\\
  Delft, Netherlands\\
  \texttt{Yanxia.Zhang@tudelft.nl}
  \and
  Hayley Hung\\
  Delft University of Technology\\
  Delft, Netherlands\\
  \texttt{h.hung@tudelft.nl}
}

\maketitle
\thispagestyle{empty}

\begin{abstract}
With the tremendous progress in sensing and IoT infrastructure, it is foreseeable that IoT systems will soon be available for commercial markets, such as in people's homes. In this paper, we present a deployment study using sensors attached to household objects to capture the resourcefulness of three individuals. The concept of resourcefulness 
highlights the ability of humans to repurpose objects spontaneously for a different use case than was initially intended. 
It is a crucial element for human health and wellbeing, which is of great interest for various aspects of HCI and design research. Traditionally, resourcefulness is captured through ethnographic practice. Ethnography can only provide sparse and often short duration observations of human experience, often relying on participants being aware of and remembering behaviours or thoughts they need to report on. Our hypothesis is that resourcefulness can also be captured through continuously monitoring objects being used in everyday life. We developed a system that can record object movement continuously and deployed them in homes of three elderly people for over two weeks. We explored the use of probabilistic topic models to analyze the collected data and identify common patterns.
\end{abstract}

\section{Introduction}
Internet of Things (IoT) systems are emerging for commercial markets, which serve as an opportunity for resourcefulness research. This paper addresses the research question of whether resourcefulness can be captured through sensor data attached to objects being used in everyday life. Using pervasive sensors and computational tools provides a novel approach for us to understand how objects are used in everyday life. Traditional methods are based on human ethnography interviews, surveys, self-report and logging. These methods can suffer from subjective bias. In addition, they can only sample limited information from experiences recalled from the users' perspective. We propose that it would be desirable to uncover patterns of behavior that would not be easy to observe through conventional ethnographic methods. By equipping continuous passive sensing to households objects, we hope to ultimately be able to discover routines such as daily or nightly, weekday or weekend patterns. This information is valuable in understanding the usage patterns of objects across practices which can be useful for suggesting new design practices.

\begin{table*}[t]
\centering
\caption{Summary of Participants and objects to track in each household}
\label{tab:object}
\begin{tabular}{ccll}%{| c | c | c{1cm} | c{1cm} |}
\hline
ID  & Age          & Housing & Objects to track \\ \hline
1  & 69  & Terraced house & Remote Control, Spider Stick, Garden Door, Fridge, Breakfast Chair, Tray \\ \hline
2  & 76      & Apartment & Chair Pillow, Remote Control, Rope on Stairs, Kitchen Drawer, Fridge \\ \hline
3  & 74 & Terraced house & Kitchen Chair, Fridge, Remote Control, Kitchen Cabinet Door, Knitting Needle, Tablet  \\ \hline
\end{tabular}
\end{table*}

To address our research questions, we propose to use machine learning and leverage sensor data to capture resourcefulness, in parallel with ethnography. Specifically, our project aims to understand resourcefulness of third-age adults (people who have generally just retired, and are still fit and able). Previous works have identified that resourcefulness emerges in daily practices \cite{KuijerNicenboimGiaccardi2017}. Hence, we focus on identifying situations where objects have been used in a creative and unexpected situation. For example, it can be when objects have been used together or at unexpected times during the day. To record events of object usage with sensors, we use their movement as a proxy of objects being handled by the participants.

Our first step is to examine raw events recorded by the sensor. These are a set of time series data that correspond to events triggered by object movement in chronological order, coming from each object and individual sensing modality. At this step, we are interested in whether the detected events from the sensors are consistent with observations during ethnographic fieldwork.
 
We further explore interactions between objects to uncover patterns of objects used together and their temporal context. To achieve this, we created feature representations to encode object movements for the entire group of objects per household. We extract common usage patterns by applying a topic model. Since topic models are probabilistic, they also generate quantitative metrics to rank the saliency of these patterns. We are interested in whether machine learning can generate meaningful patterns that are related to resourcefulness without human annotation. 

\section{Data Collection}
A set of six TI SensorTag (CC2650STK) \footnote{\url{http://www.ti.com/ww/en/wireless_connectivity/sensortag/}} were deployed at the lower most hierarchy, which collects the object usage data from the Inertial Measurement Unit (IMU) and environment sensors. The data is streamed to a Raspberry Pi, which acts as a data collection gateway at the site and regularly records the data on a remote server (IBM Cloud). The software that was loaded on the Raspberry Pi is located at the resourceful$-$gateway repository \footnote{\url{https://github.com/resourceful-ageing/resourceful-gateway}}. The software on the sensors is found at the resourceful$-$sensortag repository \footnote{\url{https://github.com/resourceful-ageing/resourceful-sensortag}}.

We initially conducted a deployment study in the homes of five elderly people for over two weeks. The initial deployment for the first two households were not successful due to technical issues such as sensor disconnection and running out of battery. Subsequently, we refined the system development and managed to collect data continuously for over two weeks in the remaining three households. The characteristics of the household data and tracked objects are summarized in Table \ref{tab:object}. The environment data (air pressure, humidity, temperature and light) is sampled every minute. The movement data from the IMU is logged using a sleep and wake up strategy, logging continuously for 20 seconds whenever the amplitude of movements is above a certain threshold. The threshold was determined through a pilot test with two researchers explicitly manipulating the sensor tags to ensure that noisy jitters are not recorded.

\section{Computational Model}
We first extracted the movement data and then applied Latent Dirichlet Allocation (LDA) \cite{BleiNgJordan2003} to discover hidden patterns. Our first step is to convert raw movement data of objects to word tokens and build a vocabulary. We consider the data from one day to be a document and each movement instance as a word. Then, we form the document collections as features represented with word distributions over the built vocabulary. Each corpus consists of data from one household during the field deployment. Finally, we use LDA to identify topics in the documents, in other words, daily usage patterns.

\subsection{LDA for identifying object interaction routine}
\begin{figure}[htp]
  \centering
    \begin{tikzpicture}[scale=0.9, transform shape]
    \tikzstyle{main}=[circle, minimum size = \gmcsize, thick, draw =black!80, node distance = \gmcoffset]
    \tikzstyle{connect}=[-latex, thick]
    \tikzstyle{box}=[rectangle, draw=black!100]
      \node[main, fill = white!100] (alpha) [label=below:$\alpha$] { };
      \node[main] (theta) [right=of alpha,label=below:$\theta$] { };
      \node[main] (z) [right=of theta,label=below:z] {};
      \node[main, fill = black!10] (w) [right=of z,label=below:w] { };
      \node[main] (beta) [right=of w,label=below:$\bm{\beta}$] { };
      \path (alpha) edge [connect] (theta)
            (theta) edge [connect] (z)
            (z) edge [connect] (w)
            (beta) edge [connect] (w);
      \node[rectangle, inner sep=0mm, fit= (z) (w),label=below right:N, xshift=13.5mm] {};
      \node[rectangle, inner sep=4.2mm,draw=black!100, fit= (z) (w)] {};
      \node[rectangle, inner sep=4.6mm, fit= (z) (w),label=below right:M, xshift=14.5mm] {};
      \node[rectangle, inner sep=9mm, draw=black!100, fit = (theta) (z) (w)] {};
    \end{tikzpicture}
  \caption{Graphical model representation of LDA model}
  \label{fig:lda_graph}
\end{figure}
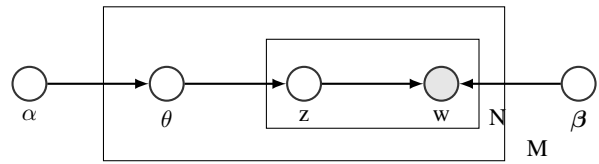

Topic models (i.e., LDA \cite{BleiNgJordan2003}) have been successfully applied in text mining community to extract summaries of large and unstructured collection of documents. LDA treats words in documents as a generative probabilistic process, in which hidden variables describe the structure of how topics are composed of mixture of words and how documents are composed of mixture of topics. Figure \ref{fig:lda_graph} illustrates the generative process of the LDA model. Assuming we have a corpus $D$ consisting of $M$ documents and $N$ words per document, LDA learns the topic proportions of documents by representing each topic as a mixture of word distribution and each document as a mixture of topic-document distribution. The probability of each topic $\theta$ is initialized with a Dirichlet prior with hyperparameter $\alpha$. The word probability over each topic is denoted as $p(w|z_{n}, \beta)$ with parameter $\beta$. 

The idea of using topic models to discover object usage routines is that we consider the combination of object movements are a mixture of events related to activities from daily practices. If we consider a breakfast routine, this would involve the handling of different objects involving the kitchen drawer, fridge and mug in the morning. On the other hand, a leisure routine could be composed of a mixture of using remote control, fridge and sofa chair at late evening. If we treat each of these routines as a topic that consists of commonly co-occurring object usage instances, we can consider each day to be a mixture of social interaction routines in the same way that a document is formed by a mixture of topics. Typically for data clustering, popular methods include the K-means clustering which hard assigns a data sample to one particular cluster, or Gaussian mixture models (GMM) which assigns a soft membership score for each cluster assuming a Gaussian density \cite{christopher2016pattern}. The benefit of LDA is that it provides a richer representation for topics that are composed of probabilistic description of words, in our case, the object movement events.

\subsection{Feature extraction and word representation}
\begin{figure}[t]
\centering
\includegraphics[width=0.99\columnwidth]{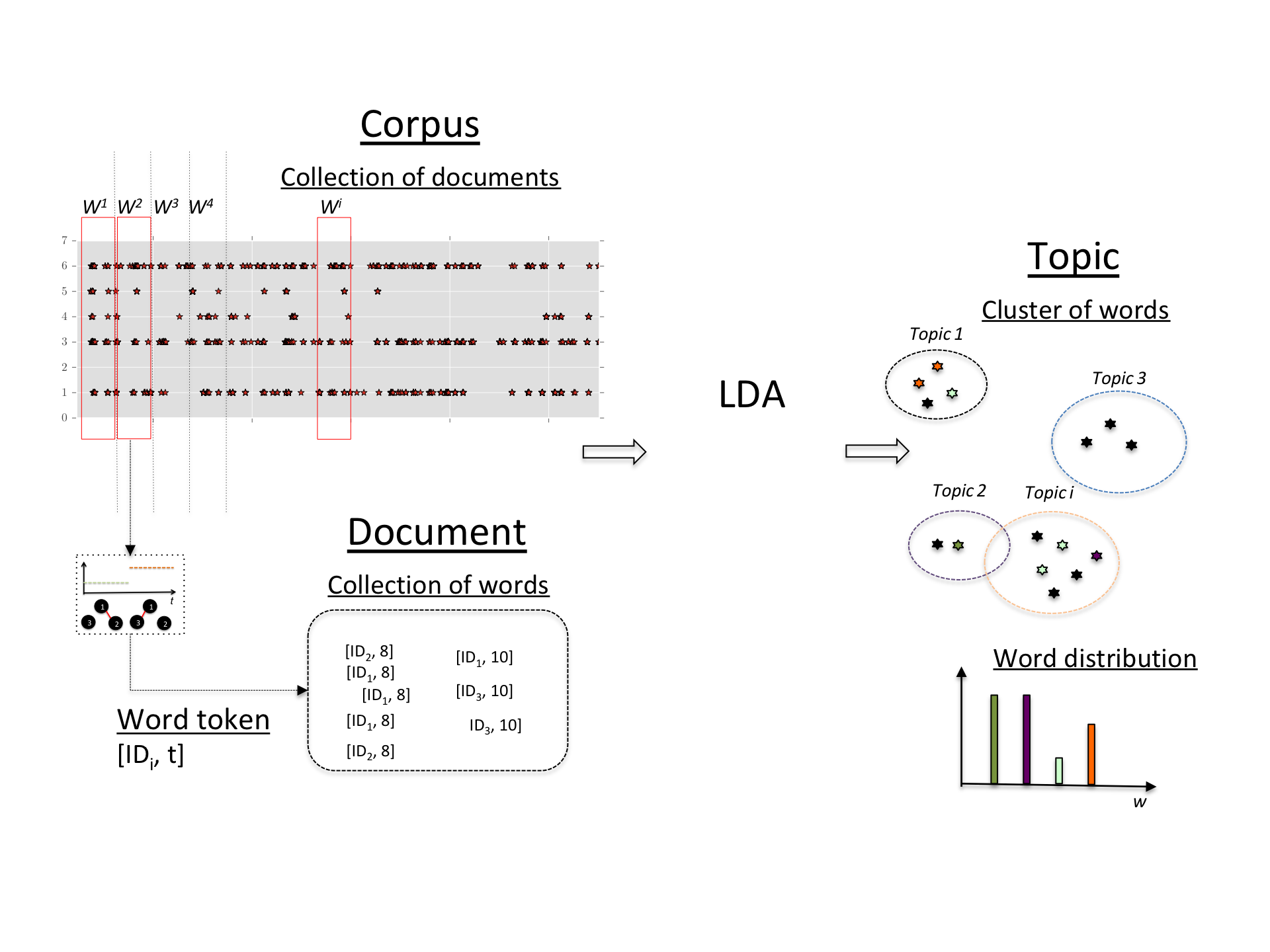}
\caption{An illustration of the workflow}
\label{fig:lda_method}
\end{figure}

We used a similar approach that have successfully applied LDA to mine human interaction patterns from infrared data \cite{ZhangOlenickChangEtAl2018}. The code of our data analysis is available at the resourceful-topicmodels repository \footnote{\url{https://github.com/resourceful-ageing/resourceful-topicmodels}}. Figure \ref{fig:lda_method} illustrates the workflow of feature extraction and word representation from the movement event data. Our first step is to convert raw infrared log data to word tokens and build a vocabulary. We consider the recorded data from one day to be a document and each object movement event as a word. Then, we form the document collections as features represented with word distributions over the built vocabulary. Finally, we use LDA to identify topics in the documents.

We constructed a corpus consisting of documents $D = \{\mathbf{w^{1}}, ..., \mathbf{w^{M}}\}$ to represent the interaction data during M days, where one day $\mathbf{w}$ is a document in the corpus. Each document $\mathbf{w}$ consists of $N$ words $\mathbf{w} = \{w_{1}, ..., w_{N}\}$ that represent interaction events within that day. The number of words $N$ can vary daily depending on the number of times an object is moved during that day.

\subsubsection{Word tokens}
As the IMU sensors were sampled at 20Hz, this generated time series of discrete observations when movement occurred at every 0.05s. If the object moves continuously, data related to movement amplitude, the ID of the object and timestamp will be continuously recorded at the predefined resolution. We encoded the word tokens with both the timestamp and ID information of the object only when movement events were detected. We used a non-overlapping windows of $\Delta_{t} = 1$ hour to extract movement instances and construct words $w$. The word $w$ for object $ID_{i}$ is represented as:
\begin{equation}
w= [ID_{i}, t]
\end{equation}
where $ID_{i}$ is the current object and $t$ is the temporal context that describes the hour of the day in a 24 hour format. In this way, we implicitly encode the movement duration as more words will be generated when more movement events observed within a period of $\Delta_{t}$.

\subsubsection{Choice of Number of Topics}
The choice of the optimal number of topics is usually selected via a cross validation procedure (see \cite{ZhangOlenickChangEtAl2018} for more details). The idea is to choose a cost metric (i.e., perplexity) and select the number that best fits the training data and also generalizes well to unseen data. This is an important step because, as with any unsupervised method, metrics are required to validate the quality of the clustering in some way. Unfortunately,  due to the limited number of days available in our dataset there are not enough data samples to learn optimal model parameters. Therefore, we opted for a heuristic approach. The 'rule of thumb' is that the number of topics should not exceed the number of documents (number of days). Since we only have two weeks of data, we choose to set the number of topics as ten in this work.

\begin{figure*}[t]
\centering
\includegraphics[width=0.98\textwidth]{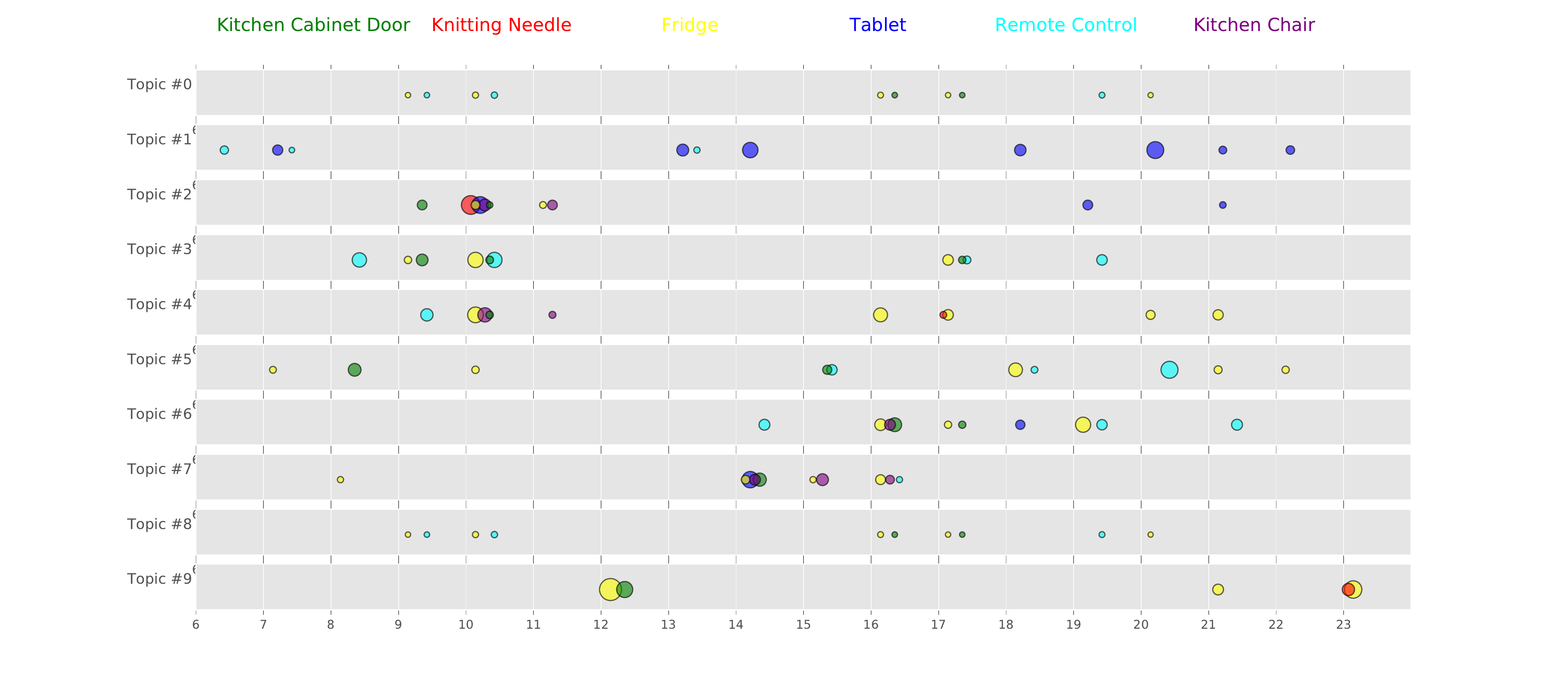}\\
\caption{Extracted topics for one household. Each row is one topic and its associated word events. The color indicates the object and the size corresponds to the probability of the movement events. The x axis shows the hour of the day.}
\label{fig:lda_topics}
\end{figure*} 

\section{Result}
To illustrate the identified patterns, we plot the extracted 10 topics from the data of household 3 in Figure \ref{fig:lda_topics} where each color corresponds to the specific object. Each topic is a distribution over words and corresponds to an object usage routine, in other words, a mixture of objects being used together. Each row is a collection of interaction event words indicated as circles. Larger circle indicates a higher probability of those interaction events occurring in that topic. For clarity of display, Figure \ref{fig:lda_topics} shows only the words with probability greater than 0.01. The ordering of the circles within each time window is arbitrary. Therefore, for a given hour of the day, multiple circles appearing in that window can only be considered as having been moved during the same hour. We refer to this in the remainder of the text as co-use. We can see that topic 0 and topic 8 are identical. Duplicated topics appearing suggests that the number of routines is less than the parameter 10 we set in this data.

\subsection{Object co-use in different routines} 
If we examine each individual topic, each of them appears to associate with different routine behaviours. For instance, topic 0 consists of a repeated pattern of two combinations of object usages including 'kitchen Cabinet Door' with 'Fridge' and 'Remote Control' with  'Fridge'. This topic is associated with routines around 'Fridge' which indicates its significance in this household's daily activities. The combination of 'Remote Control' with  'Fridge' usage at around 9:00-10:00 could be a daily morning ritual such as the participant watching TV while eating breakfast. The same combination appears at around 19:00-20:00 indicating the same routine of watching while eating also appearing in the evening. On the other hand,  the use of the 'kitchen Cabinet Door' with 'Fridge' occurs at around 16:00-17:00 which is most likely cooking behaviour, which is less surprising.

\subsection{Salient combination of object usage}
If we examine words occurring at the same 1 hour period, we can observe a variety of configurations of object co-use combinations. It is very common that one or two objects are used. These topics could also be related to resourcefulness (though we can only verify this with the participants themselves). For instance, topic 1 is mostly related to usage of digital device and shows a less unique routine that the tablet are used throughout the day. We also observe a few cases where three or more are co-used. These cases are particularly interesting as the higher number of objects involved might correspond to a complex situation where something creative or resourceful is happening with the co-use of the objects. On the other hand, it could be a random and coincidental anomaly that has no meaning for the participant. Besides looking at the number of co-used objects, we also observe unexpected combinations of object usage. For instance, topic 2 shows a combination of five objects including 'knitting Needle', 'tablet' and other objects in the kitchen. The combination is preceded by using the kitchen door and followed by two other objects (fridge and kitchen chair) again in the kitchen. One possibility could be that the participant uses the tablet to search for food menus. However, the usage of knitting needle in that situation is uncommon and is a case that might be interesting to probe the participant further during field interviews.  

This section used two examples to discuss ways to interpret the patterns by looking at shared objects and combination of objects and associate those with situations where resourcefulness might have occurred. Nonetheless, our interpretation has limitations as they are entirely based on the continuous digital traces collected from the sensors attached to the household objects. 

\section{Conclusion}
This paper explored the use of machine learning, specifically topic models, to identify resourcefulness from IoT sensors data. We conceptualize resourcefulness as the creative usage of a combination of objects. We use the movement of objects as a proxy for object usage. The topic models were proposed to extract patterns of co-usage the raw movement data. Our results show that machine learning can extract higher level routine behaviors but alone is not sufficient to identify resourcefulness. It is important to verify the associated patterns identified by the machine learning algorithm with semantic meanings in order to generate meaningful insights to understand resourcefulness. In the future, we will explore the dynamics of usage routines and their patterns at various temporal and spatial resolutions on a larger dataset. 

\section*{Acknowledgment}
This research was supported by the project 'Resourceful Ageing' funded by NWO/STW under the Research through Design program (2015/16734/STW). 

\bibliographystyle{abbrv}
\bibliography{main}
\end{document}